\documentclass[aps,prl,showpacs,nofootinbib,onecolumn,10pt]{revtex4-1}
\usepackage{color,amsmath,amssymb,graphicx,latexsym,lineno}
\usepackage{threeparttable,multirow,txfonts,subfigure,booktabs}

\usepackage{lineno}
\usepackage{multirow}
\usepackage[figuresleft]{rotating}
\usepackage{mathrsfs}
\usepackage{ulem}

\usepackage{hyperref}
\makeatletter
\def\UrlAlphabet{%
      \do\a\do\b\do\c\do\d\do\e\do\f\do\g\do\h\do\i\do\j%
      \do\k\do\l\do\m\do\n\do\o\do\p\do\q\do\r\do\s\do\t%
      \do\u\do\v\do\w\do\x\do\y\do\z\do\A\do\B\do\C\do\D%
      \do\E\do\F\do\G\do\H\do\I\do\J\do\K\do\L\do\M\do\N%
      \do\O\do\P\do\Q\do\R\do\S\do\T\do\U\do\V\do\W\do\X%
      \do\Y\do\Z}
\def\UrlDigits{\do\1\do\2\do\3\do\4\do\5\do\6\do\7\do\8\do\9\do\0}
\g@addto@macro{\UrlBreaks}{\UrlOrds}
\g@addto@macro{\UrlBreaks}{\UrlAlphabet}
\g@addto@macro{\UrlBreaks}{\UrlDigits}
\makeatother

\graphicspath{{./}{figures/}}

\def\nat{Nature\ }
\def\aap{Astron.\ Astrophys.\ }

\def\apj{Astrophys.\ J.\ }

\def\apjs{Astrophys.\ J.\ Supp.\ }

\def\mnras{Mon.\ Not.\ Roy.\ Astron.\ Soc.\ }

\def\prd{Phys.\ Rev.\ D\ }
\def\prl{Phys.\ Rev.\ Lett.\ }

\def\araa{Annu.\ Rev.\ Astron.\ Astrophys.\ }
\def\jcap{J.\ Cosmol.\ Astropart.\ Phys.\ }

\begin{document}
%\linenumbers
%setpagewiselinenumbers
%modulolinenumbers[5]
%linenumbers
%\modulolinenumbers[5]
%pagewiselinenumbers
%\switchlinenumbers

\title{
A GeV-TeV particle component and the barrier of cosmic-ray sea in the  {Central Molecular Zone}}
\author{Xiaoyuan Huang$^{1,2}$, Qiang Yuan$^{1,2}$, Yi-Zhong Fan$^{1,2}$}
\affiliation{$^1$Key Laboratory of Dark Matter and Space Astronomy, Purple Mountain 
Observatory, Chinese Academy of Sciences, Nanjing 210023, China\\
$^2$School of Astronomy and Space Science, University of Science and 
Technology of China, Hefei, Anhui 230026, China}
%\email{xyhuang@pmo.ac.cn (XH), yuanq@pmo.ac.cn (QY), yzfan@pmo.ac.cn (YZF)}

\begin{abstract}
 {Cosmic rays are important probe of a number of fundamental physical problems such as the acceleration of high and very high energy particles in extreme astrophysical environments.} The Galactic center is widely anticipated to be an important cosmic-ray 
source and the observations of some Imaging Atmospheric Cherenkov 
Telescopes did successfully reveal a component of TeV-PeV cosmic 
rays in the vicinity of the Galactic center.  {Here we report} the 
identification of GeV-TeV cosmic rays in the central molecular zone 
with the $\gamma$-ray observations of the  {Fermi Large Area Telescope}, whose spectrum and 
spatial gradient are consistent with that measured by the Imaging Atmospheric 
Cherenkov Telescopes but the corresponding cosmic-ray energy density is
substantially lower than the so-called cosmic-ray sea component, suggesting 
the presence of a high energy particle accelerator at the Galactic center 
and the existence of a barrier that can effectively suppress the penetration 
of the particles from the cosmic-ray sea to the central molecular zone.
\end{abstract}

%\pacs{96.50.S-,96.50.sb,98.70.Sa}

\maketitle

{\bf Introduction}

It is believed that in the Milky Way the cosmic rays (CRs) could be accelerated 
by shock waves in supernova remnants (e.g., \citep{2013A&ARv..21...70B}) or 
stellar winds of massive stars \citep{2019NatAs...3..561A}.
Those charged relativistic particles would then propagate diffusively in 
the Galactic magnetic field, possibly experiencing re-acceleration, 
convection, spallation, and energy loss processes 
\citep{1990cup..book.....G, 2007ARNPS..57..285S}.  Such processes would lead to a large-scale, quasi-steady-state 
CR sea, which distributes relatively smoothly in the Galaxy, as supported
by the  {Fermi Large Area Telescope (Fermi-LAT \citep{2009ApJ...697.1071A})}  observations
\citep{2015A&A...580A..90Y,2019PhRvD..99l3001S,2018arXiv181112118A}. 

However, in the proximity of a recent or currently active accelerator, 
the smoothly distributed CR sea would be overlaid with a component 
of fresh CRs. Observations of such a fresh CR component will be very
important in studying the acceleration, injection, and transportation
processes of CRs. The Galactic Center (GC) region contains a supermassive 
black hole, Sagittarius A$^\star$, and other types of particle accelerators 
such as pulsar wind nebulae and supernova remnants. The large scale 
bubbles in $\gamma$-rays \citep{2010ApJ...724.1044S}, radio
\citep{2019Natur.573..235H}, and X-rays \citep{2020Natur.588..227P}, 
and the so-called X-ray chimney \citep{2019Natur.567..347P}, may be 
consequences of energetic activities of Sagittarius A$^\star$ in the past.
The GC region was then widely regarded as a very attractive astrophysical 
laboratory for studying the cosmic ray astrophysics. 
%quite a number of physical problems, including the production
%and propagation of high-energy CRs.
 
A bright $\gamma$-ray point source at the GC, observed by some instruments 
from GeV to TeV (e.g., \citep{Aharonian:2008zza}), may indicate an episodic 
injection of high-energy particles from past 
activities of the central black hole \citep{Chernyakova:2011zz,Malyshev:2015hqa}.
With the Fermi-LAT data, a nearly symmetric and extended excess component on
top of the diffuse background (the so-called Galactic Center GeV excess; GCE \citep{2011PhLB..697..412H,2015PhRvD..91l3010Z,2015JCAP...03..038C}) was 
identified, which could be from either dark matter annihilation or a group of 
unresolved faint point sources like millisecond pulsars \citep{2020PhRvD.101b3014C}. 
In the central molecular zone (CMZ) region, the observations of High Energy 
Stereoscopic System  {(H.E.S.S. \citep{2003ICRC....5.2811H})}  and other Imaging Atmospheric Cherenkov 
Telescopes (IACTs) reveal a large amount of very-high-energy (VHE) CRs with 
a hard spectrum and a high density (a factor of $\sim10$ times higher
than that measured locally at the Earth), implying an injection of CRs by 
a source close to the GC \citep{2006Natur.439..695A,2016ApJ...821..129A, 
2017A&A...601A..33A}. Further studies \citep{2016Natur.531..476H, 
2018A&A...612A...9H,2020A&A...642A.190M} show that there is a gradient 
of the CR density profile, $\propto r^{-\alpha}$ ($\alpha \sim 1-1.2$), 
via a deconvolution of the $\gamma$-ray profile with the gas density in the CMZ.  
Likely, Sagittarius A$^\star$ was more active in the past and had accelerated 
CRs up to PeV energies which diffuse 
outwards and collide with molecular gas to produce energetic $\gamma$-rays. 
Finally, a component of $\gamma$-rays from interactions between the CR
sea and the materials in the GC region is expected to present, which is,
however, not properly addressed in some analyses. 

To reveal the nature of the \sout{new} component discovered by VHE observations, 
it would be essential to scrutinize different emission components in the CMZ 
region and to identify possible counterpart in the low-energy $\gamma$-ray 
band \citep{2014PhRvD..89f3515M,2017PhRvL.119c1101G}.
Adopting the radio or VHE $\gamma$-ray emission from the Galactic
ridge as templates, ref. \citep{2014PhRvD..89f3515M} analyzed Fermi-LAT
data and reported an emission component which was suggested to come 
from a population of nonthermal electrons in the Galactic ridge. 
Ref. \citep{2017PhRvL.119c1101G} analyzed the data in the CMZ region, 
and concluded that a large fraction of the $\gamma$-ray emission 
measured by H.E.S.S. and Fermi-LAT might originate from the interaction 
of the diffuse Galactic CR sea with the massive molecular clouds in the 
CMZ, though the possibility of an additional component cannot be ruled out. 
In such an analysis, the GCE component, which may play an important role 
in shaping the result, was not included. To critically address the CR 
population in the most central region, it is necessary to carry out a 
dedicated morphological and spectral analysis of the Fermi-LAT data in 
the GC region, with proper incorporation of the GCE component. 
This is the main goal of this work.

Here, we report the re-analysis of the Fermi-LAT data ( {see methods, subsection  The Fermi-LAT data and subsection Point sources and the diffuse model components}) in the CMZ region and the identification of a component of 
GeV-TeV CRs, which is likely the  low energy part of the \sout{new} TeV-PeV CR component 
discovered by the IACTs  {\citep{2006Natur.439..695A,2016ApJ...821..129A, 
2017A&A...601A..33A}}. This supports the presence of a high-energy particle 
accelerator at the GC. We further show that the inferred energy 
density of CRs in the CMZ region is clearly lower than that from 
an extrapolation of the CR sea distribution. A natural explanation is that 
there is a barrier surrounding the CMZ, maybe due to the strong magnetic 
field in such a region, that can effectively suppress the penetration of 
particles from outside to the CMZ.\\

%*****************************************
{\bf Results}

{\bf Reduced GeV-TeV CR density within the CMZ.}

\begin{figure}
\centering
\includegraphics[width=0.8\textwidth]{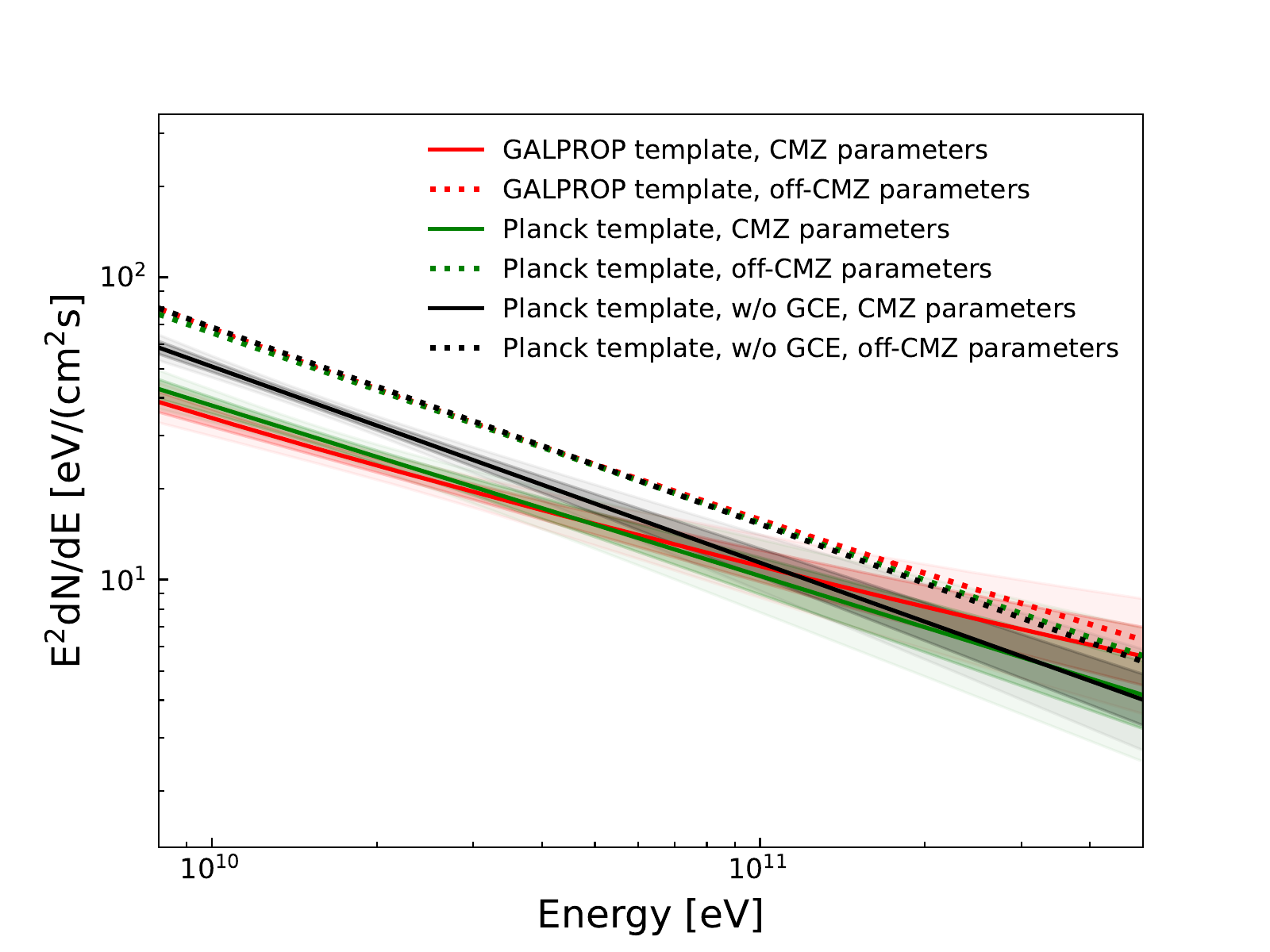}
\caption{{\bf Energy flux of $\gamma$-ray from neutral pion decays in the 
CMZ region.} Colored regions with solid lines (red and green) show the 
fitted results assuming different spatial templates, i.e., the neutral pion 
decay template from GALPROP and Planck dust opacity template in the CMZ region.
Dark and light regions are the 68\% and 95\% confidence ranges {, respectively}. 
Dotted lines show the predicted results using the off-CMZ parameters for given 
templates, which represent the anticipated CR sea component. The spectrum derived 
from the CMZ region is harder and the integrated flux is lower than those 
anticipated from the CR sea interaction. 
As a comparison, the black line and shaded band show the fitting results 
derived with the Planck template, but without subtracting the GCE component. 
Dotted black line shows the anticipated result using the off-CMZ parameters 
in this case. Again we find that the integrated flux in the CMZ is lower than 
that anticipated from the CR sea interaction.  {Source data are provided as a Source Data file.}
\label{fig:flux}}
\end{figure}

We single out the CMZ region \citep{1999ApJS..120....1T} for studying the 
properties of CRs near the GC, and the rest is referred to as the off-CMZ 
region ( {see methods, subsection The CR densities in the CMZ and off-CMZ regions}). The red line and associated bands 
in Fig.~\ref{fig:flux} show the best-fit and $1\sigma$, $2\sigma$ confidence 
regions of $\gamma$-ray fluxes from the neutral pion decay component in the 
CMZ region using the template calculated by the GALPROP code 
\citep{2000ApJ...537..763S,2011CoPhC.182.1156V}. 
Similar results have been obtained if we change the GALPROP template 
to the Planck dust opacity template \citep{2016A&A...596A.109P}, which can
avoid the uncertainty from the $X_{\rm CO}$ factor when converting the 
gas tracer emissivity to the molecular gas density, as shown by the green 
line and associated bands in Fig. \ref{fig:flux}. The spectral index of 
the CMZ region is harder by $0.19 \pm 0.07$ than that (about 2.68) predicted 
in the GALPROP Galactic diffuse emission (GDE) model A 
\citep{2015ApJ...799...86A} ( {see methods, subsection Point sources and the diffuse model components}), and the integrated 
energy flux is $(7.05 \pm 0.44)\times$ 10$^{-5}$ MeV~cm$^{-2}$~s$^{-1}$ from
8 GeV to 500 GeV. As a comparison, we show in Fig.~\ref{fig:flux}
with dotted lines the predicted spectrum in the CMZ with the parameters for the 
off-CMZ region, which is the ``expected'' emission from interactions between the 
CR sea and the gas in the CMZ assuming a relatively flat distribution of the sea. 
The spectral index of the CMZ is clearly harder than that expected from the 
interaction of CR sea with the gas, but is consistent with the VHE 
$\gamma$-ray observations (2.32 $\pm$ 0.05$_{\tt{stat}}$ $\pm$ 0.10$_{\tt{syst}}$ 
and 2.28 $\pm$ 0.03$_{\tt{stat}}$ $\pm$ 0.20$_{\tt{syst}}$
\citep{2016Natur.531..476H,2018A&A...612A...9H,2020A&A...642A.190M}). 
Furthermore, the integrated flux is lower than that predicted from the CR sea 
interaction. Our results thus suggest that there is a hard component in the 
CMZ which possibly coincides with the VHE component, and the CR density of the 
background sea component in the CMZ region should have been suppressed. 

\begin{figure*}[!htb]
\centering
\includegraphics[width=1.0\textwidth]{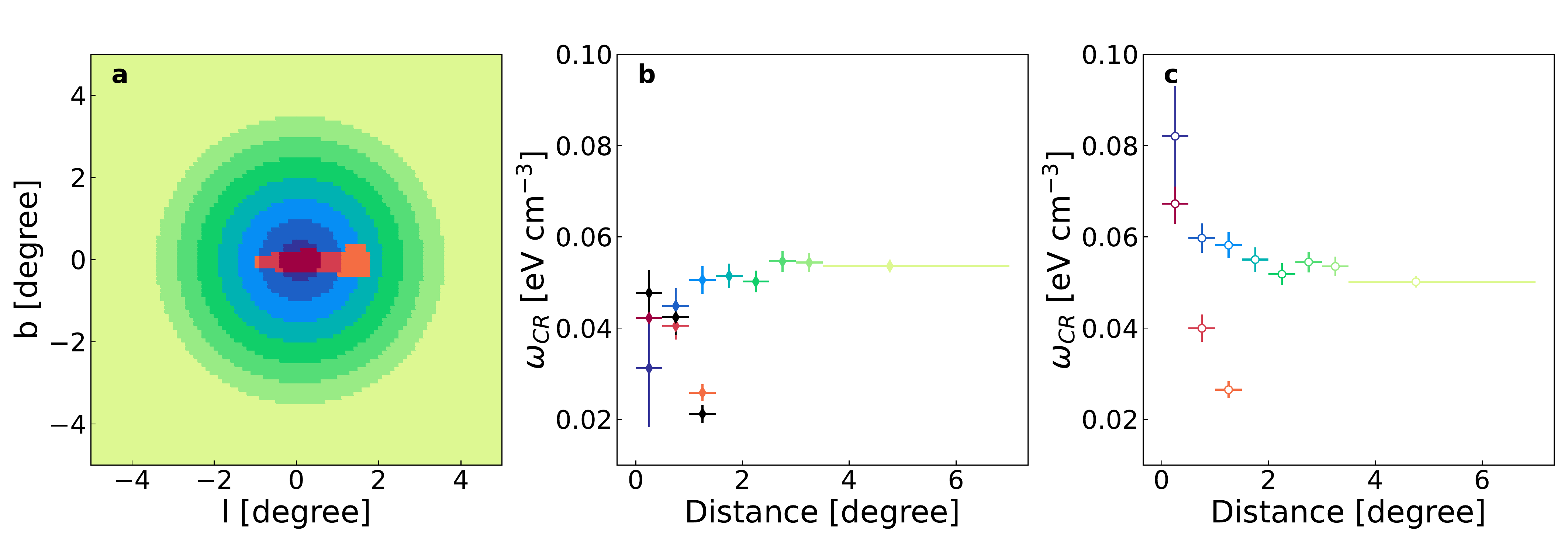}
\caption{{\bf  {The segment division of the GC region and the inferred CR  densities distribution}}.   {{\bf a} The segments, annuli centered on the GC with a width of 0.5 degrees except for the last one, where CR densities are derived}, in the CMZ and off-CMZ are marked in red to 
orange and blue to green, respectively.  {The same color code is used to show data points of CR densities in corresponding segments in {\bf b} and {\bf c}.}  {{\bf b} CR densities distribution from fittings with the GCE.} For segments outside the CMZ, the CR density is almost a constant. 
In the CMZ, the CR density declines quickly as the distance increases from the GC. 
These facts strongly suggest different physical origins of CRs within and outside the CMZ.
CR densities in the CMZ, inferred from analysis with the CS map 
\citep{1999ApJS..120....1T}, are shown with black points, which agree well with 
those derived from the fitting with the Planck map.  {The error bars represent the $1\sigma$ statistical uncertainties.}
 {\bf c}  CR densities distribution from fittings without the GCE, for which the inferred 
CR densities near the GC were boosted compared with those shown in  {\bf b}.  {The error bars represent the $1\sigma$ statistical uncertainties.  {Source data are provided as a Source Data file.}} 
\label{fig:segments_ED}}
\end{figure*}

Then we investigate the CR density distribution in the CMZ and off-CMZ
directions. To enable a high enough spatial resolution of the gas distribution,
the Planck dust opacity map is used as a tracer. We split the Planck map into 
small segments, as shown in panel (a) of Fig.~\ref{fig:segments_ED}, with a 
width of 0.5 degrees except for the last one. 
We fit the normalizations of $\gamma$-ray fluxes for each segment, and derive 
the CR  energy density $\omega_{\rm CR} = w_{\rm CR}(80~\mathrm{GeV} \leq E \leq 5 ~\mathrm{TeV})$ 
( {see methods, subsection The CR densities in the CMZ and off-CMZ regions}), as shown in panel (b) of
Fig.~\ref{fig:segments_ED}. It is very interesting to find that, for segments 
outside the CMZ, the CR  density is almost a constant, as anticipated in the 
CR sea scenario. The CR  density in the CMZ, however, declines clearly with 
the increase of distance to the GC, resembling the gradient profile observed 
at the VHE band.  The CR density in the CMZ is also generally lower than that 
outside the CMZ, confirming the results shown in Fig.~\ref{fig:flux}. 
The spatial distribution of the CR densities is different from the expectation 
that the smoothly distributed CR sea would be overlaid with a component of 
fresh CRs. This may indicate that the contribution of the CR sea has been 
suppressed in the CMZ, and the emission inside the CMZ mainly comes from 
the GeV-TeV particles associating with the VHE CR component.

{\bf The GeV-TeV counterpart of the VHE CR component.}

\begin{figure}
\centering
\includegraphics[width=0.8\textwidth]{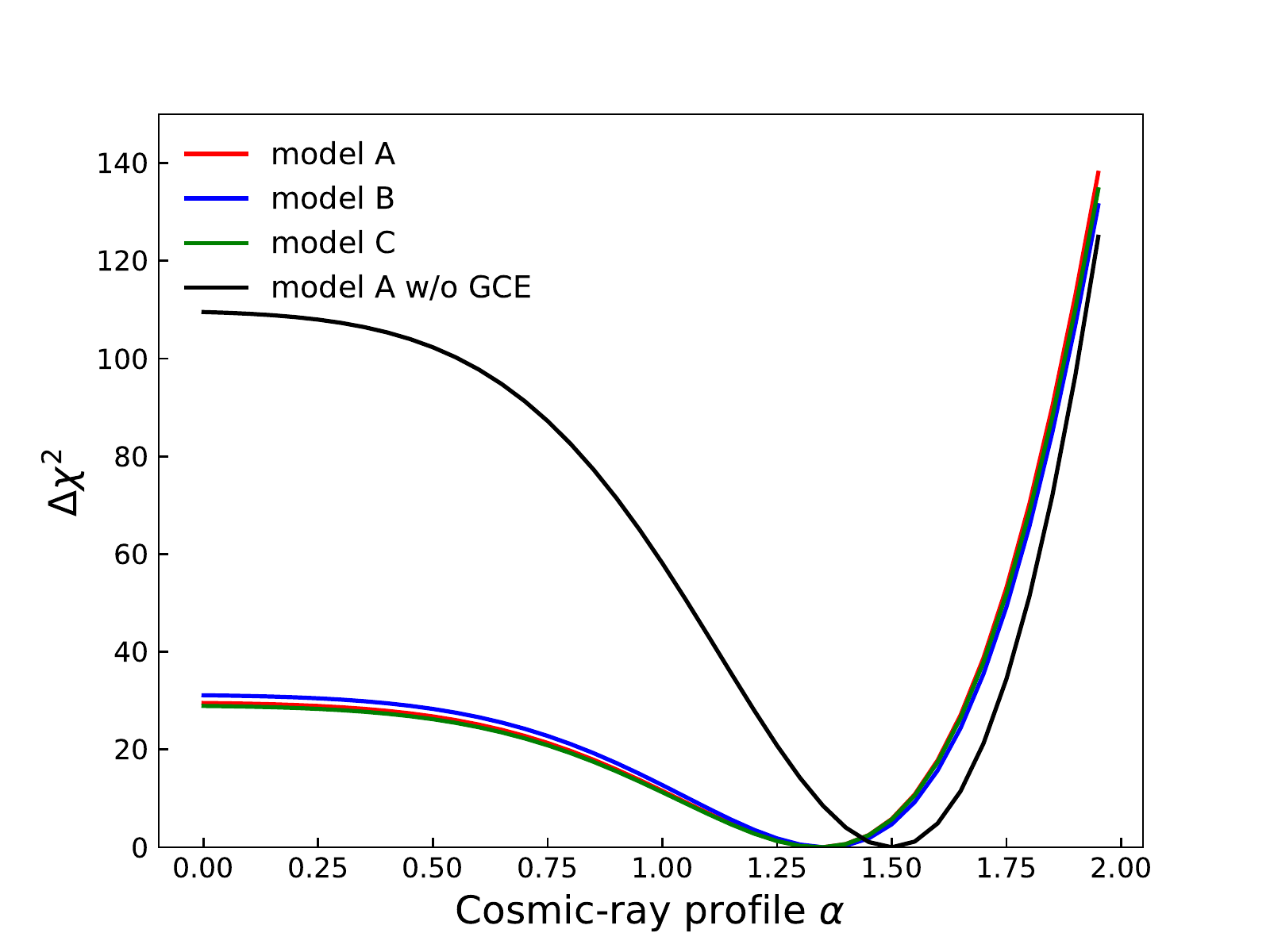}
\caption{{\bf $\Delta \chi^2$ as a function of the power-law index of the CR density 
profile, $\alpha$.} The smaller the $\Delta \chi^2$, the better the fitting. 
Results, corresponding to the GDE models A, B, and C from 
ref. \citep{2015ApJ...799...86A}, with different assumptions on the source distributions, 
the diffusion coefficients, and the re-acceleration parameters of CRs ( {see Table 2 of ref. \citep{2015ApJ...799...86A} for more details about different GDE model A, B and C}), for the off-CMZ 
region, are shown in red, blue, and green solid lines, respectively. The black solid 
line is for the fitting using the GDE model A without the inclusion of the GCE.  {Source data are provided as a Source Data file.}
\label{fig:alpha_scan}}
\end{figure}

The VHE observations of the  {hard} component in the GC show a power-law 
radial profile, $\propto r^{-\alpha}$, of the CR density with 
$\alpha=1.10\pm0.12$ \citep{2016Natur.531..476H} and $1.2\pm0.3$ 
\citep{2020A&A...642A.190M}. In the GALPROP GDE model, the spatial 
gradient of CR densities in the very center region of the Galaxy 
is not well modelled because of its limited spatial resolution. 
Furthermore, the molecular hydrogen component incorporated in GALPROP 
is traced with the CO emission, which may be strongly contaminated by 
the foreground and background emission in the GC. The CS radio map 
\citep{1999ApJS..120....1T} is believed to be a better tracer of the 
molecular component \citep{2016Natur.531..476H,2020A&A...642A.190M}.
Therefore we instead use the CS map in the CMZ region to construct the neutral
pion decay $\gamma$-ray template. We take a power-law function $r^{-\alpha}$ 
to parameterize the radial distribution of the CR density in the CMZ (ignoring
the contribution from the CR sea component) and multiply its line-of-sight 
integral with the CS emission template to model the spatial distribution 
of the hard component (referred to as the CS $\times$ $r^{-\alpha}$ 
template hereafter;  {see methods, subsection The hard CR component in
the CMZ} ). 
We scan $\alpha$ from 0.0 to 2.0 and obtained the 
difference of logarithmic likelihood values, defined as 
$\Delta\chi^2=2\ln{\mathcal L}_{\rm max}-2\ln{\mathcal L}_{\alpha,\rm max}$,
where ${\mathcal L}_{\rm max}$ is the maximum likelihood of the fitting
with varying $\alpha$ and ${\mathcal L}_{\alpha,\rm max}$ is the maximum 
likelihood of the fitting with fixed $\alpha$. The $\Delta\chi^2$ value
as a function of $\alpha$ is shown in Fig.~\ref{fig:alpha_scan}, where 
a smaller value means a better fitting.
We find that the Fermi-LAT data favors a non-zero value of $\alpha$ at a 
confidence level of 5.4$\sigma$. The inferred $\alpha=1.35^{+0.06}_{-0.09}$ 
is consistent with those derived in the VHE band 
\cite{2016Natur.531..476H,2020A&A...642A.190M}. 
To visualize the radial dependence of the CR density and to investigate 
the dependence on the mass estimation with different observations, we 
derive the CR density in the CMZ region with the CS map divided into 
3 segments, as done previously for the Planck map. 
Following ref. \citep{1999ApJS..120....1T} we estimate the mass for 
each segment with the corresponding CS flux. As shown in panel (b) of 
Figure \ref{fig:segments_ED}, the results agree well with those 
incorporating the Planck map.

\begin{figure}
\centering
\includegraphics[width=0.8\textwidth]{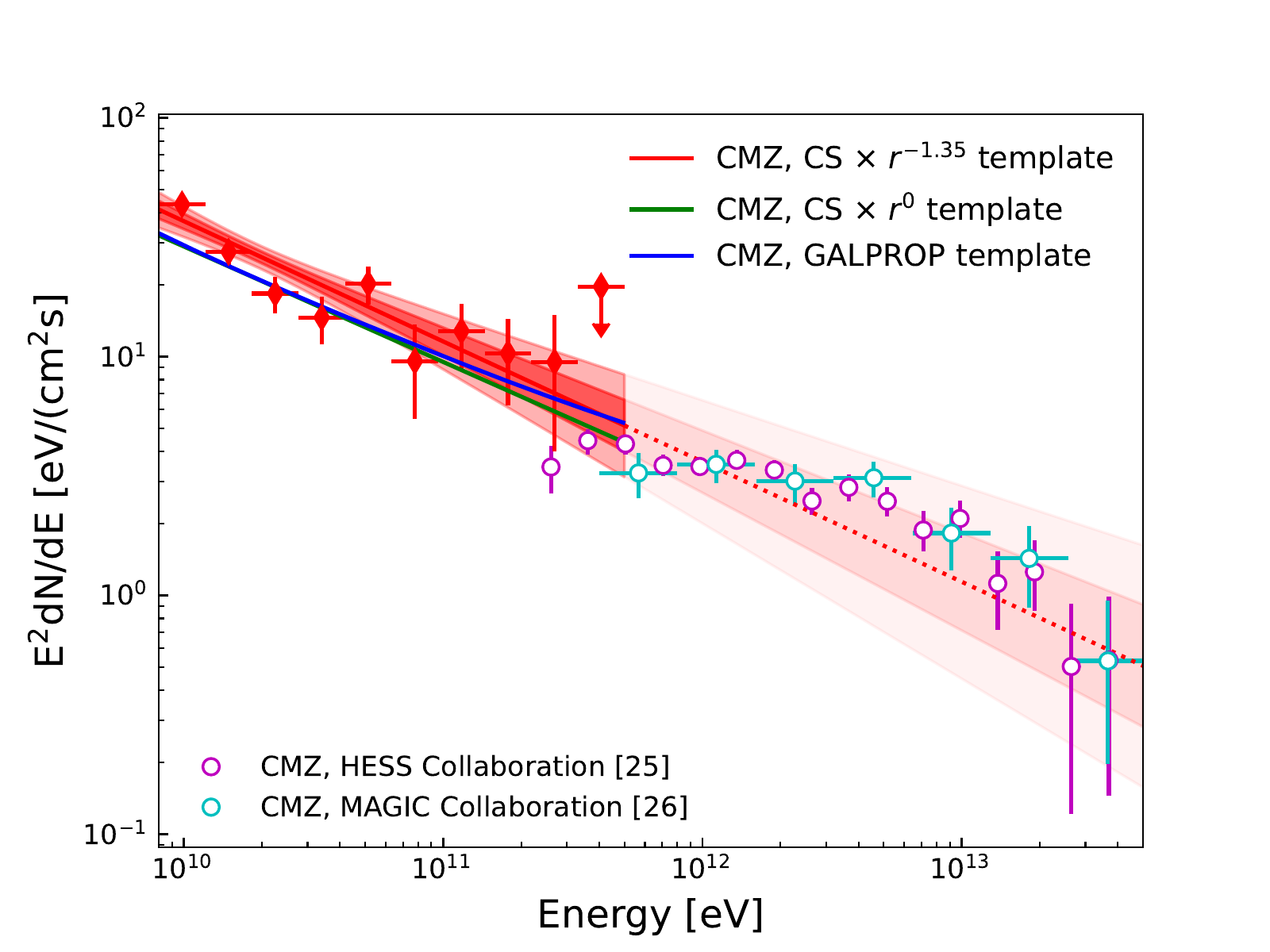}
\caption{{ {\bf Spectral energy distribution of $\gamma-$ray emission contributed by the CMZ component.}  {Red filled dots show results obtained in this work, open circles in magenta show 
measurement by H.E.S.S. \citep{2018A&A...612A...9H} and open circles in cyan show measurement by MAGIC \citep{2020A&A...642A.190M}.}} The  {red}  solid line and associated  {dark and light} bands show the best-fit power-law 
model and the 68\% and 95\% confidence ranges,  {respectively, for the CS $\times$ $r^{-1.35}$ template}.  {The error bars represent the $1\sigma$ statistical uncertainties and the upper limit is at the 95\% confidence limit.}  {Best-fit power-law models for the CS $\times$ $r^{0}$ template and  the GALPROP H$_2$ template are also shown in green and blue solid lines as a comparison.} 
The lighter red band and red dotted line above 500 GeV are extrapolations from 
the low energy fitting for the model with CS $\times$ $r^{-1.35}$ template.  {Source data are provided as a Source Data file.}
\label{fig:flux_sed}}
\end{figure}

We derive the spectral energy distributions (SED) of the hard component
in the CMZ region, using the CS $\times$ $r^{-1.35}$ template, and report
the results in Fig. \ref{fig:flux_sed}. The spectral index of the 
$\gamma$-rays in the energy range from 8 GeV to 500 GeV is $2.50 \pm 0.08$, 
which is roughly consistent with that measured in the VHE band 
(2.32 $\pm$ 0.05$_{\tt{stat}}$ $\pm$ 0.10$_{\tt{syst}}$ and 
2.28 $\pm$ 0.03$_{\tt{stat}}$ $\pm$ 0.20$_{\tt{syst}}$ by H.E.S.S.
\citep{2016Natur.531..476H, 2018A&A...612A...9H}). 
As a test of systematics, we also fit the data with the CS $\times$ $r^{0}$
template and the GALPROP H$_2$ template to investigate the dependence of the 
CMZ CR spectrum on the spatial templates, and always get a hard spectrum.
The fitting qualities of these two templates, however, are significantly 
poorer than that of the CS $\times$ $r^{-1.35}$ template.
For the CS $\times$ $r^{0}$ (GALPROP H$_2$) template we have $-2\ln({\mathcal L})=
105826.6$ (105831.4), while for the CS $\times$ $r^{-1.35}$ template we have 
$-2\ln({\mathcal L})=105797.2$ (see also Fig. \ref{fig:alpha_scan} for the differences of the log-likelihood values for different density 
profile slope $\alpha$). The $\gamma$-ray flux between 8 and 500 GeV in the 
inner 150 pc ($\pm 1^\circ$) region is $(5.50 \pm 0.38)\times 10^{-5}$ 
MeV~cm$^{-2}$~s$^{-1}$. Taking the mass of dense molecular clouds as $3\times10^{7}$ 
M$_{\odot}$ \citep{1999ApJS..120....1T,2007A&A...467..611F} and the nuclear 
enhancement factor $\eta_{\rm N}=1.5$ for heavier nuclei ( {see methods, subsection The CR densities in the CMZ and off-CMZ regions}), 
we have $\omega_{\rm CR} = (4.5\pm0.3)
\times10^{-2}$ eV~cm$^{-3}$ in the CMZ, which is close to the local CR density of 
$\sim 4.3\times10^{-2}$ eV~cm$^{-3}$ in the same energy range \citep{Zyla:2020zbs}. 
The proton number density above 10 GeV,  which is $(4.3\pm0.3)\times10^{-12}$
cm$^{-3}$ supposing the power-law spectrum holds for energies down to 10 GeV,
is about 3 times smaller than the expectation of the CR sea in the GC region \citep{2016ApJS..223...26A}. Note that the derived proton number density in the CMZ 
region in ref. \citep{2016ApJS..223...26A} is also lower than that from the conventional 
model prediction, which supports our results.\\

%*********************************
{\bf Discussion}

\begin{figure}
\centering
\includegraphics[width=0.8\textwidth]{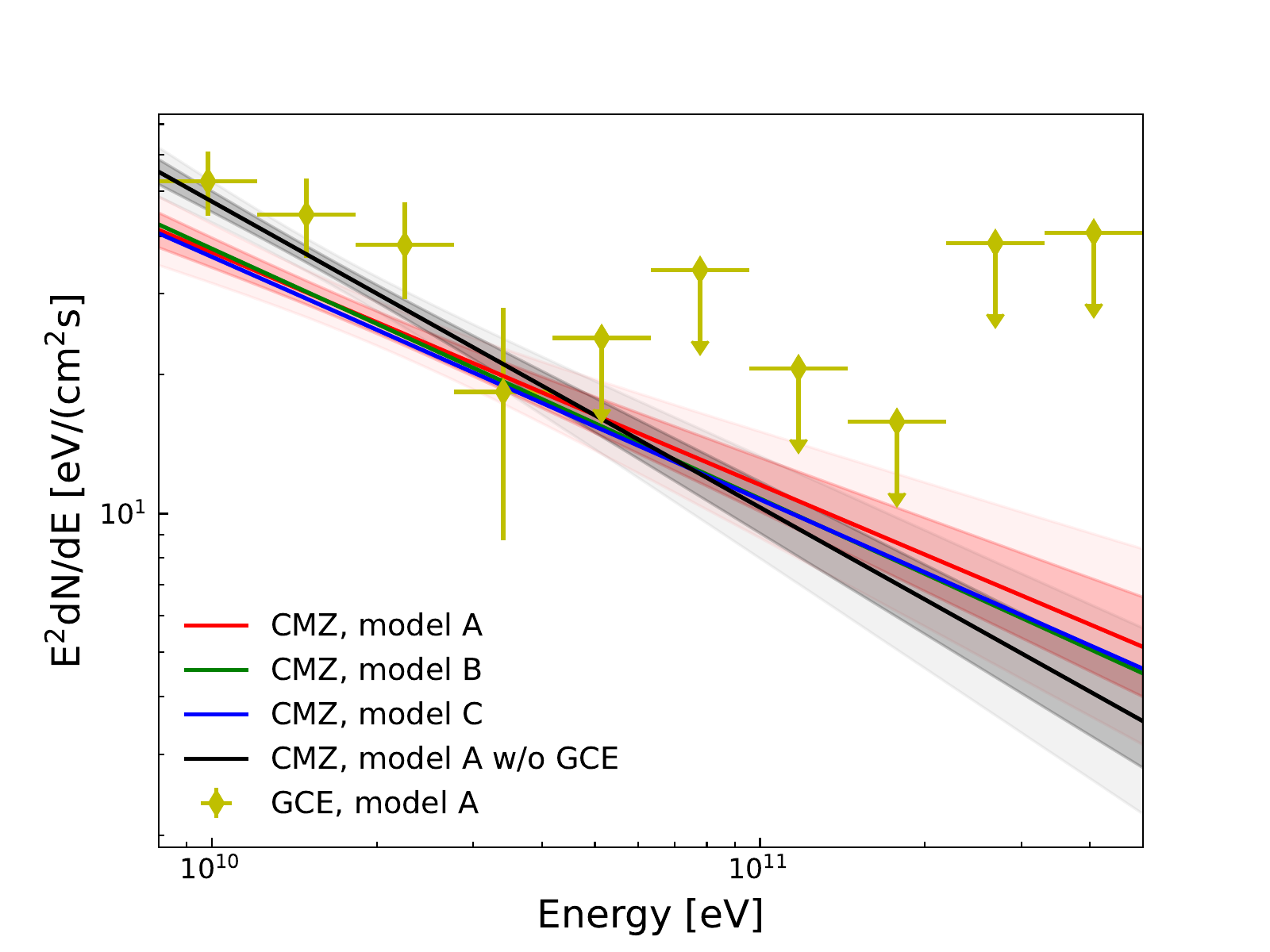}
\caption{{ {\bf Energy fluxes of $\gamma$-ray emission from  neutral pion decays associated with H$_2$ in the CMZ region.}  {To test the robustness of the results, different GDE models of the other emission 
components, such as the templates for emissions from the inverse Compton 
scattering or the neutral pion decay in the off-CMZ region, are used in the fitting. With the CS $\times$ $r^{-1.35}$ template, the red solid line and associated dark and light bands are for results using GDE model A. The blue and green solid lines are for results using GDE model B and GDE model C, respectively.} The black band and
line are for the fitting using GDE model A but without incorporating the GCE}. 
The SED of the GCE component is also shown in yellow points for reference. 
Dark and light regions show the 68\% and 95\% confidence ranges of the power-law 
fitting spectrum,  {respectively}.  {The error bars represent the $1\sigma$ statistical uncertainties and the upper limits are at the 95\% confidence level.  {Source data are provided as a Source Data file.}}
\label{fig:CSGC}}
\end{figure}

The systematics from the choice of the GDE model could be important in 
affecting the results, as found in previous studies \citep{2012ApJ...750....3A,
2015PhRvD..91l3010Z,2015JCAP...03..038C,2015ApJ...799...86A}.
Here we use several GDE models, namely models A, B, and C from 
ref. \citep{2015ApJ...799...86A}, with different assumptions on the source 
distributions, the diffusion coefficients, and the re-acceleration 
parameters of CRs ( {see Table 2 of ref. \citep{2015ApJ...799...86A} for more details about different GDE model A, B and C}), to test the robustness of the results. 
Fig.~\ref{fig:CSGC} shows the spectra of the hard component 
derived assuming the CS $\times r^{-1.35}$ template, for different GDE 
models of the other emission components. The results show good agreement 
with each other. Also we scan $\alpha$ for these different GDE models, 
and find again very similar results (see Fig. \ref{fig:alpha_scan}). 

The GCE could contribute to the diffuse $\gamma$-ray emission in the CMZ region. 
It has an intensity profile of $r^{-2\beta}$ with $\beta \sim 1.1-1.4$ \citep{2015JCAP...03..038C,2016PDU....12....1D}. 
The spectrum of the GCE component is soft above a few GeV, as shown in 
Fig. \ref{fig:CSGC} which is derived assuming a $r^{-2.56}$ 
profile (a line-of-sight projection is further applied when fitting
the data). In the VHE band, the effect of GCE on the  {hard} component should 
be negligible. At GeV energies, however, the flux of the GCE is higher 
than that of the hard component, and may affect the analysis significantly. 
Without addressing the GCE component in the analysis, we get a softer spectrum
with an index of $2.64 \pm 0.06$ ($2.65 \pm 0.07$) for the CMZ
region using a Planck template (CS $\times$ $r^{-1.35}$ template), 
as shown in Fig.\ref{fig:flux} (Fig.\ref{fig:CSGC}). 
This is consistent with the inferred spectral index of CR protons in
ref. \citep{2018arXiv181112118A}, $2.80\pm0.03$, considering the fact that the
energy-dependence of the inelastic interaction cross section would result 
in a hardening of $\sim0.1$ for $\gamma$-rays compared with CRs.
We also compare the proton density at 10 GeV in the CMZ with previous work. 
In our analysis, it is $(1.00\pm0.05) \times 10^{-12}$ GeV$^{-1}$~cm$^{-3}$, 
which is close to the value for Sgr B complex in \citep{2018arXiv181112118A}.
Note that besides the CMZ region, the spectrum of the point source, 
Sagittarius A$^\star$, should also be influenced by the contamination 
from the GCE which is beyond the focus of this work. 
The impact of the GCE on the CR density distribution in the inner Galaxy region 
can be seen in panel (c) of Fig. \ref{fig:segments_ED}. We can see that
the CR density still declines quickly with increasing distance from the GC in the 
CMZ region, consistent with ref. \citep{2017PhRvL.119c1101G}, although in which the
comparison between the CMZ and off-CMZ CR densities was not carried out. 
The background model without the GCE will increase the CR density in the innermost 
part of the GC region. For sky segments far away from the innermost GC, we can still 
observe a flux decrease from the off-CMZ region to the CMZ region, implying the 
barrier effect of the CR penetration. The impact of the GCE on the density profile 
of the hard CMZ component is shown in Fig. \ref{fig:alpha_scan}. The best-fit 
slope parameter $\alpha$ becomes steeper, $\sim1.5$, if the GCE is not subtracted. 

As found in the above analysis, the Fermi-LAT observations
indicate that the contribution of the CR sea, which might be 
accelerated by sources distributed in the Galactic plane and diffuse
across the whole Galaxy, should be suppressed in the CMZ. However,
the $\gamma$-ray fluxes shown in Fig.~\ref{fig:flux} or the CR 
densities in Fig.~\ref{fig:segments_ED} reflect the line-of-sight
integrals of those quantities.  
To estimate the impacts due to the projection effect and also 
the uncertainties of the foreground and background modeling, we adopt the 
quasi-three-dimensional gas distributions used in the GALPROP package 
and apply randomly scaling factors to the gas annuli (except for the 
innermost annulus which is our target), which are in the face-on 
view of the Galactic plane and centered on the GC, based on the fitted 
distribution of the normalization factor, and re-fit the density contrasts 
of the off-CMZ region to the CMZ region.  {Please see method, subsection The foreground/background effect for the detailed 
description.} We find that the CR density contrast is always larger than 2
for the segment with distance to the GC between 1.0 and 1.5 degrees, 
consistent with that shown in Fig.~\ref{fig:segments_ED}. We conclude 
that the CMZ should indeed play a role to block CRs from entering into 
the very center region.

Several mechanisms can impede CR penetration into molecular clouds, such as the
effect of magnetic field compression and the self-excited magnetohydrodynamics 
turbulence \citep{2015ARA&A..53..199G, 2018ApJ...855...23I,2018ApJ...868..114D}.
Taking an analogy of the solar modulation effect where low energy CRs are blocked 
outside of the solar system by the magnetic field associated with solar winds, 
we briefly discuss the suppression of CR penetration in the CMZ. Employing the 
solar modulation model of ref. \citep{1971ApJ...167..169O} as an example, we assume a 
spherical region of the CMZ with a radius of $R=200$ pc and replace the solar wind 
velocity by the galactic wind velocity. The velocity of the galactic wind is not 
certain in the GC region \citep{2011MNRAS.413..763C,2017PhRvD..95b3001T, 2018AdSpR..62.2731A,2019arXiv191211057Y}, and we take it $V_0\sim500$ km~s$^{-1}$ 
as a reference. We assume that the diffusion coefficient $D_0$ is constant in 
the CMZ, and neglects its energy-dependence. Note that this is different from 
ref. \citep{1971ApJ...167..169O} in which the diffusion coefficient was assumed to be 
proportional to radius $r$. We have tested that assuming the same radius-dependence of the diffusion coefficient as ref. \citep{1971ApJ...167..169O}, the same scaling relation as obtained analytically can be reproduced. We solve numerically the transportation equation of particles
\begin{equation}
\frac{1}{r^2}\frac{\partial}{\partial r}\left(r^2V_0S-
r^2D_0\frac{\partial S}{\partial r}\right)=\frac{qV_0S}{r},
\end{equation}
where $S(r)$ is the spatial part of the particle distribution
which we separate from the energy part, $q\approx-1$ is related to
the energy spectrum of particles. Adopting $D_0=3\times10^{28}$
cm$^2$~s$^{-1}$ (Note that $D_0$ depends on the chosen 
value of $V_0$ and is for CR particles with energy of several hundreds of GeV),
we can obtain a scaling relation of $\sim r^{0.8}$
for the CR density distribution for $50<r/{\rm pc}<200$ (corresponding to 
$0.3^{\circ}-1.3^{\circ}$ angular range), which means the CR density decreases 
by a factor of $\sim 3$ from $r=200$ pc to $r=50$ pc. Therefore the CR fluxes for 
$r<200$ pc can be efficiently suppressed.  A  {hard} CR component  with a density 
profile of $\sim r^{-1.35}$ has been revealed by the Fermi-LAT data, which is 
consistent with that shown by the VHE observations. Taking the same framework 
of the CR transportation as discussed above, and assuming a steady point-like 
source at the GC, we obtain an approximate $r^{-1.5}$ profile of the CR density
for $50<r/{\rm pc}<200$, which is close to the observed value. Note that
the pure diffusion predicts a $r^{-1}$ profile, and the convection results 
in a steeper profile. The Fermi-LAT data may imply a deviation from the $r^{-1}$ 
distribution, which could be a hint of the existence of convective winds. 
Considering that the diffusion coefficient should be energy-dependent, and at 
higher energies diffusion may be more important than convection, we may expect 
that the CR distribution in higher energy band is closer to $r^{-1}$ than in lower 
energies, which is just consistent with the observations. Alternatively, 
a time-dependent particle injection may also lead to a steeper radial profile. 
Detailed modeling is necessary, which is beyond the scope of the current work.

We note that the current analysis is probably limited by the projection 
effect of three-dimensional distributions of both the gas density and 
the CR density \citep{2020A&A...642A.190M}. More precise three-dimensional 
gas models, particularly in the innermost region of the GC, may further improve
our understanding of the CR origin as well as the transportation in the GC
\citep{2007A&A...467..611F}.

%*****************************************
{\bf Methods}

{\bf The Fermi-LAT data.} 

The Fermi-LAT data 
of version P8R3 and class SOURCE are used in this analysis. 
We select the data recorded from August 4, 2008 to February 1, 2020, 
in total 600 weeks. Since one of our goals is to investigate the spatial
distribution of the CR density in the GC region, where crowded point 
sources may affect the morphology of the diffuse emission, data with 
good angular resolution are crucial in this analysis. In addition,  
the GCE which peaks around few GeV \citep{2015PhRvD..91l3010Z, 2015JCAP...03..038C} may
also significantly affect the low-energy analysis. We therefore select
photons with energies higher than 8 GeV, which have a balance of a good 
angular resolution and a sufficiently high statistics. To suppress the contamination 
from $\gamma$-ray generated by CR interactions in the upper atmosphere, 
photons collected at zenith angles larger than 90$^\circ$ are removed. 
Moreover, we adopt the specifications (DATA$\_$QUAL$>$0) $\&\&$
(LAT$\_$CONFIG==1) to select good quality data. We bin the data, 
from 8 GeV to 500 GeV, into 20 logarithmically distributed energy bins 
and $100\times100$ spatial bins with pixel size of 0.1$^\circ$ centered
at the GC. We employ the binned likelihood analysis method to analyze 
the data with the Fermitools { version 1.2.1}. The instrument 
response function (IRF) adopted is  P8R3\_SOURCE\_V2.

{\bf Point sources and the diffuse model components.}

The source model XML file is generated using the user contributed tool 
 {make4FGLxml.py} 
based on the 4FGL catalog \citep{2020ApJS..247...33A,2020arXiv200511208B}.
The spectrum of Sagittarius A$^\star$ is modeled as a power-law instead
of the default log-parabola spectrum. Two additional point sources, 
3FHL J1747.2-2822 and 3FHL J1748.6-2816 which are not in the 4FGL source 
catalog {{gll\_psc\_v23.xml}} 
but in the 3FHL source catalog \citep{2017ApJS..232...18A} {{gll\_psch\_v12.xml}}, 
are also added in our analysis. Power-law spectra of the two 3FHL sources are
assumed.

We use the GDE modeled by the GALPROP code in this work, which 
gives different components of the diffuse emission individually. 
The result of model A as introduced in ref. \citep{2015ApJ...799...86A} is
taken as our baseline GDE model. We also adopt their models B and C,
which assume different CR source distributions and diffusion coefficient,
for cross check. For systematics studies about the template tracing the
neutral pion decay $\gamma$-rays, besides the mentioned GALPROP generated 
model, we also employ the Planck dust opacity map from 
ref. \citep{2016A&A...596A.109P} as an alternative, which traces the gas 
distributions without the uncertainty from the $X_{\rm CO}$ factor of 
the conventional molecular gas tracer.
The bremsstrahlung in our energy range is expected to be small and is thus neglected. The $\gamma$-ray emissions from the inverse Compton 
scattering of high energy electrons off the optical, infrared, and the
cosmic microwave background are taken as another template. The third 
diffuse template is the isotropic background. Power-law spectra of 
these three diffuse templates are assumed, and both the normalizations 
and spectral indices are treated as free parameters in the likelihood fit. 
For the GCE component, we use a line-of-sight integration of $\rho^2(r)$,
where $\rho(r)\propto r^{-1.28}$ as derived in ref. \citep{2015JCAP...03..038C}, 
to model its spatial distribution. Again we assume a power-law to model
the spectrum of the GCE. As shown in ref. \citep{2015JCAP...03..038C}, the power-law 
spectrum is reasonable to describe the high-energy tail of the GCE. 

To summarize, the model used in the likelihood fitting includes: point sources
from the 4FGL and 3FHL catalogs, the inverse Compton scattering component of the GDE, 
the isotropic diffuse background, the GCE, and the neutral pion decay component of the GDE, 
with the last one being our target and various spatial templates are used during the 
analysis. To show in detail the $\gamma$-ray components used in our analysis, we list 
all the different components which are adopted in the fittings in  {a table supplied as Supplementary Table 1}.

{\bf The CR densities in the CMZ and off-CMZ regions.}

To obtain the CR density in the CMZ region, the spatial template tracing the
neutral pion decay is split into the CMZ region and off-CMZ region. The spectral 
indices and normalizations of the CMZ and off-CMZ regions are fitted independently.

To further reveal the CR density distribution without pre-defined spatial 
template, we split the Planck template into segments as defined in 
Fig. \ref{fig:segments_ED}. We model each segment with a power-law spectrum 
by fixing the spectral index to be 2.56 (2.64) in the CMZ (off-CMZ) region, 
i.e., the values derived from the previous fittings with CMZ and off-CMZ 
division. We fix all the parameters of the point sources outside the CMZ to 
their best-fitting values as well. The CR densities ($w_{\rm CR}$ in eV/cm$^3$) 
in those segments can thus be estimated from the normalizations (and hence
luminosities) of the $\gamma$-ray emission as
\begin{eqnarray}
  w_{\rm CR}(E_{\rm CR}) && \approx  0.018 
  \left(\frac{\eta_{\rm N}}{1.5}\right)^{-1}\nonumber\\
  &&\times \left(\frac{L_\gamma(E_\gamma)}{10^{34}~\mathrm{erg/s}}\right)
  \left(\frac{M}{10^6 M_{\odot}}\right)^{-1}\text{,}
\end{eqnarray}
where $E_{\rm CR} \approx 10 E_{\gamma}$ is the corresponding energy of 
CRs giving $\gamma$-ray energy of $E_{\gamma}$, $\eta_{\rm N}$ accounts for the
correction from nuclei heavier than protons and is taken as 1.5 in this work, 
$L_\gamma(E_\gamma)$ is the $\gamma$-ray luminosity, and $M$ is the total mass 
of the gas in the segment which is estimated using the relation between 
the dust opacity and the column density \citep{Planck:2011ah}.

{\bf The hard CR component  in the CMZ.}

In order to compare with the VHE measurements consistently, we use the 
CS map multiplied the projected radial profile of the CR density to model the 
$\gamma$-ray emission from the CR interaction with the molecular gas in the CMZ 
region. For the neutral pion decay template generated by GALPROP, we mask the 
contribution from H$_2$ traced by CO in the CMZ to avoid double counting it. 
The contributions from atomic and ionized hydrogen, which are expected to be 
sub-dominant, are kept as foreground in the GDE model. Note that the background 
CR sea may also contribute to the $\gamma$-ray emission in the CMZ even if most 
of it is blocked due to the barrier effect of the CMZ. This may add up to the 
hard CR component. We neglect it in the current treatment. Also the hard CR 
component should extend to the off-CMZ region. Due to its fast decline, we do
not consider it outside the CMZ region. 

{\bf The foreground/background effect.} 

To examine the effect due to the foreground and background, we use the 
outputs of GALPROP predicted templates calculated from the Galactocentric 
annuli of gas as an approximation of a three-dimensional modeling. We rebin 
the gas file into 6 annuli, which are in the face-on view of the Galactic 
plane and centered on the GC, with radii of [0, 1.5) kpc, [1,5, 3.5) kpc, 
[3.5, 5.5) kpc, [5.5, 8) kpc, [8, 10) kpc, and [10, 50] kpc, respectively. 
We take the first segment, from 0 to 1.5 kpc, as our target, and split it 
further into the CMZ region and off-CMZ region. 
In the fitting yielding Fig. \ref{fig:flux} with the GALPROP template, 
the normalization and index for the off-CMZ component, which is the same for
all annuli, are fitted to be $1.19\pm 0.03$ and $2.63\pm 0.04$, respectively.
As an approximation, for the five annuli except the innermost one, we fix their
spectral index to be $2.63$, but randomly assign the normalization parameters 
according to a Gaussian distribution with central value of 1.19 and standard 
deviation of 0.15 (5 times larger than that obtained in the fitting). 
All other components (three point sources in the CMZ, the inverse Compton
scattering GDE, the isotropic diffuse emission, the GCE, and the CMZ and
off-CMZ regions of the innermost annulus) are left free to fit. 
With this setup, we expect that the projection effect from the region outside 
the innermost annulus will be reduced. We further divide the innermost annulus
into segments as that shown in panel (a) of Fig.~\ref{fig:segments_ED}, and
fit the normalizations of each segment. We find that the CR density ratios of 
the off-CMZ to the CMZ segment for distance bin from 1.0 to 1.5 degrees to the
GC are mostly larger than 2, confirming the finding that the CR density in the 
CMZ is lower than that outside the CMZ. However, we should also note that CO 
emission, instead of CS emission, is used to trace the distribution of molecular 
hydrogen near the GC in the GALPROP, and this may lead to unaccounted for systematics.

{\bf Data Availability}\\ The Fermi-LAT observation data analyzed/used in this work are publicly available  {(\url{https://heasarc.gsfc.nasa.gov/FTP/fermi/data/lat/weekly/photon/}).  {The Fermi-LAT 4FGL catalog could be found at \url{https://fermi.gsfc.nasa.gov/ssc/data/access/lat/10yr\_catalog/} and the Fermi-LAT 3FHL catalog could be found at \url{https://fermi.gsfc.nasa.gov/ssc/data/access/lat/3FHL/}. The Planck dust opacity map is available at \url{https://irsa.ipac.caltech.edu/data/Planck/release_2/all-sky-maps/maps/component-maps/foregrounds/COM_CompMap_Dust-GNILC-Model-Opacity_2048_R2.01.fits}. The CS map is available at \url{https://www.nro.nao.ac.jp/~nro45mrt/html/data/tsuboi/GC.CS10.10.FITS.tar.gz}. The model maps of DGE model A, B and C could be found at \url{http://www-glast.stanford.edu/pub_data/845/}.  Source data are provided with this paper. The data that support the findings of this study are available from the corresponding author upon reasonable request. }
}\\\

{\bf Code Availability}\\ The Fermitools used in this work are publicly available  {(https://fermi.gsfc.nasa.gov/ssc/data/analysis/software/). GALPROP is available at \url{http://galprop.stanford.edu/}. The user contributed tool make4FGLxml.py for generating source model XML file is available at \url{https://fermi.gsfc.nasa.gov/ssc/data/analysis/user/make4FGLxml.py}.}

%\bibliographystyle{apsrev}
%\bibliography{sample63}

%{\bf Data availability.} The Fermi-LAT observation data analysed/used in this work are publicly available.\\

{\bf Acknowledgements}\\
We acknowledge the use of the Fermi-LAT data provided by the Fermi Science Support Center. 
We thank Zhaoqiang Shen and Bei Zhou for useful discussion.
This work is supported by the National Natural Science Foundation of China (Nos. 11921003, 
U1738205, U1738210), the Key Research Program of the Chinese Academy of Sciences (No. XDPB15),
Chinese Academy of Sciences, and the Program for Innovative Talents and Entrepreneur in Jiangsu.\\\
%the National Key Research and Development Program (No. 2016YFA0400200) 

{\bf Author Contributions}\\
X.H carried out the data analysis.  {X.H,  Q.Y and Y.-Z.F} interpreted the data and prepared the manuscript. 
\\\

 {{\bf Corresponding authors}\\
Correspondence to Xiaoyuan Huang or Qiang Yuan or Yi-Zhong Fan.}
\\\

{\bf Competing Interests}\\
The authors declare no competing interests.\\\

\end{document}